# Directed Energy Transfer from Monolayer WS$_2$ to NIR Emitting PbS-CdS Quantum Dots

Arelo O.A Tanoh[1,2], Nicolas Gauriot[1], Géraud Delport[1], James Xiao[1], Raj Pandya[1], Joo Young Sung[1], Jesse Allardice[1], Zhaojun Li[1], Cyan A. Williams [2,3], Alan Baldwin[1], Samuel D. Stranks[1,4], Akshay Rao[1*]

[1]Cavendish Laboratory, Cambridge, JJ Thomson Avenue, CB3 0HE, Cambridge, United Kingdom

[2]Cambridge Graphene Centre, University of Cambridge, 9 JJ Thomson Avenue, Cambridge, CB3 0FA, Cambridge, United Kingdom

[3]Department of Chemistry, University of Cambridge, Lensfield Rd, CB2 1EW, Cambridge, United Kingdom

[4]Department of Chemical Engineering & Biotechnology Department of Chemical Engineering & Biotechnology, University of Cambridge, Philippa Fawcett Drive, Cambridge CB3 0AS, UK.

## Abstract

Heterostructures of two-dimensional (2D) transition metal dichalcogenides (TMDs) and inorganic semiconducting zero-dimensional (0D) quantum dots (QDs) offer unique charge and energy transfer pathways which could form the basis of novel optoelectronic devices. To date, most has focused on charge transfer and energy transfer from QDs to TMDs, i.e. from 0D to 2D. Here, we present a study of the energy transfer process from a 2D to 0D material, specifically exploring energy transfer from monolayer tungsten disulphide (WS$_2$) to near infrared (NIR) emitting lead sulphide-cadmium sulphide (PbS-CdS) QDs. The high absorption cross section of WS$_2$ in the visible region combined with the potentially high photoluminescence (PL) efficiency of PbS QD systems, make this an interesting donor-acceptor system that can effectively use the WS$_2$ as an antenna and the QD as a tuneable emitter, in this case downshifting the emission energy over hundreds of meV. We study the energy transfer process using photoluminescence excitation (PLE) and PL microscopy, and show that 58% of the QD PL arises due to energy transfer from the WS$_2$. Time resolved photoluminescence (TRPL) microscopy studies show that the energy transfer process is faster than the intrinsic PL quenching by trap states in the WS$_2$, thus allowing for efficient energy transfer. Our results establish that QDs could be used as tuneable and high PL efficiency emitters to modify the emission properties of TMDs. Such TMD/QD heterostructures could have applications in light emitting technologies, artificial light harvesting systems or be used to read out the state of TMD devices optically in various logic and computing applications

**Introduction**

Monolayer transition metal dichalcogenides (TMDs), which are derived from their layered bulk crystals via dry mechanical cleavage[1] or liquid phase exfoliation[2,3] have attracted a great deal of research interest due to their unique optical, electronic and catalytic properties[4,5,6]. Monolayer TMDs can also be obtained via epitaxial growth methods, in particular chemical vapour deposition[7,8] (CVD), which is an area of ongoing research. A number of monolayer TMDs such as tungsten disulphide ($WS_2$) have a direct optical gap[5]. This property compounded with high absorption coefficients, high carrier mobilities[5] and potentially high photoluminescence quantum efficiency[9,10,11] (PLQE) promise great potential for their application in optoelectronic devices namely photodetectors, light emitting diodes (LEDs) and photovoltaics (PV)[12]. The reduced dielectric screening in the monolayer limit compared to their bulk counterparts gives rise to tightly bound electron-hole pairs (i.e. excitons) with binding energies of the order of hundreds of meV at room temperature[13,14]. As a consequence, monolayer TMDs provide a convenient medium to study diverse excitonic species that arise via exciton-exciton or exciton-charge interaction[13,15,16,17]. Alternatively, these tightly bound excitons can be funnelled to other fluorescent media where they recombine radiatively at lower energy, thus tuning the emission properties of TMD excitons. Nanocrystal quantum dots (QDs), for example, provide a convenient, colour tuneable high PLQE emission medium[18,19] to which transferred 2D TMD excitons might be funnelled.

The exciton funnelling i.e. nonradiative energy transfer (ET) process can occur via two main mechanisms, namely Förster resonance energy transfer[20] (FRET) and Dexter energy transfer[21] (DET). FRET is a long-range process (~1-11 nm) [20] that occurs via dipole-dipole coupling, where the electromagnetic near-field of an oscillating transition dipole in the donor induces a transition dipole in the acceptor. Consequently, FRET between donor and acceptor systems is dependent on their physical separation and to a large extent, the overlap of emission and absorption spectra[20,21,22]. On the other hand, DET involves direct simultaneous tunnelling of electron hole pairs from the donor to acceptor due to donor-acceptor charge orbital overlap. As such, DET is strongly distance dependent and requires extremely close proximity between donor and acceptor molecules (≤ 1nm)[21,23].

A considerable amount of research into 2D-QD heterostructures has focused on interfacial charge transfer (CT) between QDs and monolayer TMDs for applications in photodetectors[24–31] and phototransistors[32,33]. To date, studies on energy transfer in 2D-QD heterostructures for light harvesting and light sensing applications have mainly focused on 0D-2D exciton transfer where monolayer TMDs or graphene are used as efficient exciton sinks to which optically or electrically generated excitons from QD emitters are non-radiatively transferred[22,30,34–39].

Here, we demonstrate for the first time efficient ET from 2D TMDs to 0D QDs. We present a down–shifting heterostructure system, where monolayer tungsten disulphide ($WS_2$) acts as an antenna from which optically generated excitons are funnelled to lower energy lead sulphide-cadmium sulphide (PbS-CdS) near infrared (NIR) QD emitters. Photoluminescence excitation (PLE) studies confirm 2D-0D ET. Probing the underlying photophysics via time resolved optical microscopy reveals a fast, non-radiative ET process that out-competes intrinsic exciton trapping in monolayer $WS_2$. These results establish ET from 2D TMDs to 0D QDs as an efficient means to control excitonic behaviour, allowing for tuning of emission energies and construction of artificial light-harvesting systems.

**Results & Discussion**

Figure 1.a (1-6) shows the sample fabrication process from the initial exfoliated monolayers to the heterostructure. Following mechanical exfoliation of monolayer $WS_2$, a 1,3 benzendithiol self-assembly monolayer (SAM) was deposited as per *ref 40* [40]. *Following this* a single layer of oleic acid capped PbS-CdS QDs was spin-coated onto the monolayer. The choice of dithiol ligand guarantees strong adhesion of the QDs to the TMD monolayer surface. Sample preparation is detailed further in the *experimental methods* section provided in section 1 of the supplementary information (SI). Figure 1.b. illustrates the process of exciting the 2D material with high energy visible photons forming excitons that funnel to the QDs where they recombine and emit lower energy NIR photons.

Figure 1.c. shows the absorption and PL spectra of a $WS_2$ monolayer. The absorption spectrum of $WS_2$ monolayer (light blue circles) clearly reveals `A', `B' and `C' excitonic peaks positioned at 2.0 eV (617 nm), 2.4 eV (512 nm) and 2.88 eV (430 nm) respectively. The PL spectrum (dark blue dashed line) is well overlapped with the A exciton band. The absorption and PL spectra of the colloidal QDs are plotted in Figure 1.b. The colloidal PbS-CdS absorption spectrum (solid black line) reveals an absorption peak at 1.76 eV (704 nm) while the PL spectrum (black dotted line) exhibits the red-shifted peak position at 1.38 eV (900 nm). Interestingly and importantly, the $WS_2$ PL lies within PbS-CdS absorption spectrum, which is a key ingredient to open the possibility of efficient FRET. Consequently, we carefully chose PbS-CdS QDs and $WS_2$ monolayer as an efficient energy transfer pair. The steady state confocal PL spectra of QD film on the bare substrate (black) and the heterostructure (red) are plotted in Figure 1.e. While the QD film on the bare substrate shows a broad Gaussian PL peak in the NIR region centred at 1.38 eV (900 nm), the heterostructure exhibits two distinctive PL peaks i.e. the narrow $WS_2$ PL peak in the visible region centred at 2.0 eV (~619 nm) and a broad QD PL peak in the NIR region at 1.42 eV (870 nm). We note that the QD PL spectrum of the heterostructure is blue-shifted by 30 nm and enhanced by a factor of 2.6. The blue-shift in heterostructure's QD PL spectrum can be attributed to a difference in the QD's dielectric environment on the $WS_2$ monolayer compared with the bare substrate. A difference in QD aggregation concentration on the $WS_2$ monolayer compared with the QD film on the bare substrate could also contribute to the observed blue shift in QD emission. We consider that the enhancement in QD PL indicates the possibility of efficient ET of $WS_2$ excitons to the QD[21].

Figure 2a, shows the optical micrograph (left) of a $WS_2$ flake and confocal NIR PL map (right) from the same region obtained upon excitation at 514.5 nm. Enhanced NIR PL from QDs is obtained in the vicinity of the monolayer (dashed line) whereas QD PL in the bulk flakes (solid line) is quenched. The difference in NIR PL intensity between monolayer and bulk flakes suggests that the $WS_2$ monolayer serves as the ET donor, while the bulk quenches excitons. To delve into the possibility of ET from the $WS_2$ monolayer to PbS-CdS QDs we employ wide-field photoluminescence excitation (PLE) microscopy. We recorded the PL intensity integrated over the NIR region (800-1000 nm), exclusively corresponding to PL from the QDs, and scanning the excitation wavelength across 560-680 nm, mainly resonant to $WS_2$ at low fluence (c.a., ~0.006 µJ/cm$^2$ at 620 nm). PLE spectra were taken on the heterostructure (red) and in an area with QDs only (black) away from the heterostructure. As shown in Figure 2.b, unlike the PLE spectrum of QD only area (black), the PLE spectrum of the heterostructure (red) clearly reveals the signature `A' excitonic peak centred at 616 nm (~2.0 eV), indicative of a significant contribution from the $WS_2$. Furthermore, as shown in Figure 2.c., the resulting PLE spectrum (red line) obtained by subtracting the normalised QD PLE spectrum (Figure 2.b., black) from that of heterostructure (Figure 2.b., red) is almost perfectly overlapped with a typical WS$_2$ absorption spectrum (blue circles). This is strong evidence that energy transfers from WS$_2$ monolayer to the QDs. In order to quantitatively analyse the contribution of ET from WS$_2$ monolayer to QD, we calculated $PL_{ctr}$ as a function of excitation wavelength. Details on the derivation of $PL_{ctr}$ are given in SI section 2. As presented in Figure 2.d., the contribution is maximized at 616 nm with a value of 58% and reduces considerably thereafter at lower energy excitation energy. Additionally, we carried out PLE measurement on a series of heterostructures with various QD-2D surface attachment thiol ligands. As well as the heterostructure based on 1,3 benzenedithiol (BDT) reported in herein, 1,4 butanedithiol (BuDT) and 1,6 hexanedithiol (HDT) were also studied. SI section 3.1 provides a brief PLE study of the heterostructures based on the different ligands. From this we note here that all heterostructures measured show ET from 2D to QD.

All surface attachment ligands used are of lengths < 1 nm, and thus in principle lie within range for ET via tunnelling i.e. DET. Although orbital overlap between the monolayer TMD donor and QD acceptor is a possibility at such separation distances, their respective large oscillator strengths highly favours ET via FRET[21] over DET. In SI section 3.2, we estimated the theoretical Förster radius of **$R_0 \approx 6.5$ nm**, which exceeds the lengths of the ligands used. This result emphasizes the significance of the combined oscillator strengths of the constituent heterostructure materials (i.e. TMD donor and QD acceptor) over their physical separation, even at low proximity, which strongly suggests FRET as the dominant ET mechanism observed in the heterostructures measured. In addition, while short ligands such as BDT have previously been shown to improve CT between QDs[41], the CdS shell encapsulating the PbS core has been shown to suppress CT[42].

To gain further insight into the dynamics of the ET process observed from PLE we turn to time resolved PL (TRPL) microscopy, where we detect changes in emission decay from WS$_2$ using a 509 nm pulsed laser excitation. Excitation is filtered from the detection line with a 510 nm long pass filter, while QD emission is removed using a 700 nm short-pass filter, allowing for WS$_2$ monolayer PL detection only. To distinguish bare WS$_2$ from WS$_2$ in the heterostructure, we refer to the former as `pristine' WS$_2$.

Figure 3.a shows the normalized time resolved PL decay signals of the pristine monolayer and heterostructure under low fluence excitation (0.01 µJ cm$^{-2}$). The transient PL profile of pristine WS$_2$ shows bi-exponential decay components, whereas the fast component of the heterostructure's PL profile is quenched below the detector's initial response function (IRF). The two PL decay components observed in the pristine monolayer can be attributed to direct band-edge to ground state excitons transitions and exciton trapping respectively[11]. In contrast, the much faster PL kinetics observed in the heterostructure suggests an additional efficient fast decaying process present in this system. In fact, this quenching observed in the heterostructure is in accordance with what is expected of the PL dynamics of the donor in a nonradiative ET system. Figure 3.b. shows an excitation fluence series performed on both pristine and heterostructure samples. The pristine case shows a general increase in PL lifetime with fluence, which is indicative of `trap' or `defect' state filling. This trap limited behaviour has also been observed in WS$_2$ and MoS$_2$ monolayers treated with bis(trifluoromethane)sulfonimide (TFSI) [43,11]. The apparent increase in the fast component of the PL lifetime with fluence is due to trapping and detrapping of excitons to the band edge prior to recombination to the ground state. The long-lived component is due to radiative transitions from the trap to ground state[43]. Increasing the excitation fluence would lead to saturation of trap states, forbidding further trapping and promoting dominant band-edge to ground state recombination. The fluence series presented in Figure 3.b however lies below trap-state saturation, and subsequent exciton-exciton annihilation as per the increasing fast component PL lifetimes as a function of fluence. Otherwise, trap-state saturation, would be signalled by the onset of exciton-exciton annihilation, whereby the fast PL component would start to reduce as a function of excitation fluence. Interestingly, in the heterostructure case, fast PL components throughout the series are quenched below the IRF. This outcome suggests that ET rate is faster than the intrinsic exciton trapping in monolayer $WS_2$, which occurs on a time scale of few picoseconds[44,43].

Steady state PL measurements provide information on the spectral changes that occur in the $WS_2$ monolayer PL from pristine to heterostructure case. Also, comparing steady state PL with TRPL data at similar excitation intensity provides better understanding of exciton recombination pathways in the heterostructure. Figure 4.a shows a scatter plot of $WS_2$ PL integrals with their corresponding spectral position obtained from PL maps of the monolayer in pristine (blue) and heterostructure (red) cases. Maps were measured with 514 nm continuous wave (CW) laser excitation at 80.2 W cm$^{-2}$ intensity for good signal to noise ratio. It is known that different types of excitons exist in atomically thin nanomaterials, i.e., $WS_2$ monolayer. Accordingly, it is of importance to understand how different types of excitons behave and contribute differently when ET occurs. We begin with analysing steady state PL spectra as it gives an indication of the types of excitons present. Figure 4.b. shows the PL spectra of an exemplary point on the monolayer in pristine (blue) and heterostructure (red) form. The spectra were deconvoluted with Gaussian peaks which represent the neutral exciton (NE) and lower energy species ($X_2$) such as trions, which are characterized by broad low energy features in monolayer TMD spectra[11]. $X_2$ may also arise from eventual radiative recombination of neutral excitons trapped in sub-gap states. Upon recombination to the ground state, these excitons can bind with electrons to form trions, which is known to occur in n-type TMDs such as $WS_2$[11,45]. Figure 4.c shows the fitted time resolved PL of pristine (blue) and heterostructure (red) cases at high excitation intensity (3.2 µJ cm$^{-2}$ → 63.4 W cm$^{-2}$). Figure 4.d. shows the proposed radiative exciton recombination pathways resulting from the high intensity PL/TRPL comparison. Table 1 shows the fitted PL lifetimes ($\tau$) of pristine and heterostructure samples at low and high intensity excitation and ET efficiencies. ET efficiencies were computed via *equation 1*. SI section 4 provides the full derivation of *equation 1*. Heterostructure lifetimes are denoted by an apostrophe. Given that the fast component of the heterostructure's $WS_2$ PL lifetime ($\tau_1'$) is limited by the IRF, the fitted values presented in table 1 represent an upper bound.

**Table 1**: Fitted PL lifetimes of pristine and heterostructure samples and resulting estimates for ET efficiencies. Fast components of $WS_2$ PL decay in heterostructure $\tau_1'$ and transfer efficiencies $\eta_{ET}$ represent upper and lower bound values respectively due to limitations in instrument sensitivity. High intensity excitation values used for comparison with steady stat PL are italicised.

| Intensity | Pristine $\tau_1$ | Heterostructure $\tau_1'$ | Pristine $\tau_2$ | Heterostructure $\tau_2'$ | $\eta_{ET}$ |
|---|---|---|---|---|---|
| 0.21 W cm$^{-2}$ | 0.456 ns | 0.26 ns | 3.63 ns | 3.64 ns | 42% |
| *63.4 W cm$^{-2}$* | *0.62 ns* | *0.26 ns* | *2.95 ns* | *2.9 ns* | *58%* |

Statistical analysis of the scatter data in Figure 4.a. reveals an average PL quenching $\Delta PL_{AVE}$ 50% and spectral blue shift $\Delta\lambda_{AVE}$ of 7 nm from the pristine to the heterostructure case. The spectra in Figure 4.b shows that the *NE* component quenches by 50%, while $X_2$ quenches by 76%. An overall quenching of 67% was computed from the raw spectra. The large $X_2$ quenching helps to explain the spectral narrowing in the red signal and the general blue shift in Figure 4.a. Interestingly, the difference in quenching between the *NE* and $X_2$ species leaves 26% of quenched excitons unaccounted for. This implies an additional exciton recombination pathway. As $X_2$ may arise from slow exciton recombination from trap states, the excess quenching of $X_2$ excitons could be explained as non-radiative trap-QD transfer. Table 1 however reveals that the slow decay component ($\tau_2$) associated with trap-ground state transition remains practically unchanged between the pristine and heterostructure case for a given excitation intensity, i.e. $\tau_2 \sim \tau_2'$. WS$_2$ trap state to QD exciton transfer requires that $\tau_2' < \tau_2$ and therefore negates this possibility. This suggests that the excess quenched excitons may dissipate via some other non-radiative pathway.

On the other hand, table 1 shows that fast component of the bi-exponential decay associated with neutral exciton recombination[11] is quenched by 58% from $\tau_1 \sim 0.62$ ns in the pristine monolayer to $\tau_1' \sim 0.26$ ns in the heterostructure case. This lies in close agreement to the 50% *NE* quenching estimated in steady state PL. The strong fast PL decay lifetime quenching shows that ET occurs via neutral excitons transitioning from the WS$_2$ band edge to the QD acceptor, while intrinsic exciton trapping in the donor and non-radiative losses compete with this process. We therefore compute the lower bound ET efficiencies shown in table 1 using fast decay components ($\tau_1$) via *equation 1*. As previously highlighted, exciton trapping and detrapping in the donor gives rise to increasing $\tau_1$ as a function of fluence which manifests as an apparent increase in $\eta_{ET}$ as a function of fluence. While non-radiative pathways are yet to be uncovered, passivating trap states to improve donor PLQE should lead to more efficient ET from the WS$_2$ donor band edge to the QD acceptor.

$$\eta_{ET} = 1 - \frac{\tau_1'}{\tau_1} \qquad (1)$$

Figure 4.d provides a clear illustration of radiative exciton pathways in pristine (LHS) and heterostructure (RHS) cases, which is derived from the PL/TRPL comparison in Figure 4.b-c and supported by the TRPL fluence series in Figure 3.b. In pristine WS$_2$, upon excitation from the ground state a proportion of excitons instantaneously transition from the band edge to trap states on the order of few picoseconds[43] at trapping rate $k_{TR}$, while others recombine radiatively from the band edge to ground state at the rate $k_D$. Those excitons that are trapped in sub-gap states radiatively recombine to the ground state over long periods of the order of ns[43] at rate $k_2$. In the heterostructure, excitons preferentially transfer from the WS$_2$ band edge to the QD at rate $k_{ET}$, such that $k_{ET} > k_{TR}$, thus quenching the fast component $\tau_1$ lifetime below the IRF. This also explains the sizeable quenching of $X_2$ in the steady state PL spectra as there are fewer excitons being trapped in the presence of an acceptor QD. Band edge excitons that are not trapped, transferred or lost via some other non-radiative process, recombine radiatively to the ground state at $k_D$ over 10s – 100s of picoseconds[44], which is below the instrument response. The remaining emission from direct band edge recombination as shown in Figure 4.b, strongly suggests that the 2D-QD transfer pathway becomes saturated. As with trap states,

the QD band edge can become saturated, forbidding further incoming excitons, which may return to the WS$_2$ band-edge and radiatively recombine or dissipate via a non-radiative process as suggested by the `lost' quenched excitons identified from Figure 4.b.

To summarise the results from optical measurements presented, PLE studies confirm ET from monolayer 2D WS$_2$ to 0D QDs. Further PLE on heterostructures with differing surface attachment thiol ligands show ET. While all ligands lengths used lie within tunnelling distances favourable for DET (< 1 nm), the large oscillator strengths of the 2D TMD donor and QD acceptor favour FRET as given by the large theoretical Förster radius computed. The CdS shell surrounding the PbS core in the QDs provides an additional tunnelling barrier, thus supporting FRET as the dominant ET process observed. Time resolved PL studies further confirm non-radiative ET by virtue of strong quenching of donor WS$_2$ PL in the presence of the acceptor QDs. TRPL studies also strongly indicate that this transfer process is faster than intrinsic early time trapping of excitons in the WS$_2$ monolayer, which would otherwise lead to radiative or non-radiative exciton recombination via trap states in the pristine monolayer. Comparing high excitation intensity PL and TRPL measurements provides a clearer understanding of radiative recombination pathways for excitons in the TMD-QD heterostructure. The comparison implies that intrinsic exciton trapping in the TMD monolayer and a non-radiative process compete with ET from 2D to QD. Further analysis also suggests that the exciton transfer channel can become saturated at high excitation intensities.

**Conclusions**

In conclusion, we have demonstrated the ability to transfer excitons from monolayer WS$_2$ to NIR PbS-CdS QD emitters. PLE studies provide confirmation of ET, with 58 % of QD PL donated by monolayer WS$_2$. The large oscillator strengths of the donor TMD and acceptor QD lead to a large Förster radius, suggesting FRET as the dominant ET mechanism. TRPL studies reveal that the ET process is faster than intrinsic exciton trapping in monolayer WS$_2$. A comparative study between high excitation steady state PL and TRPL confirms exciton transfer from the WS$_2$ band edge to the PbS-CdS band edge, while intrinsic exciton trapping in the donor and other non-radiative channels act as competing pathways. Residual emission from the donor in the heterostructure suggests that the ET pathway can be saturated at high excitation intensities. Future studies of such heterostructures could provide a clearer understanding of non-radiative loss mechanisms via more sensitive methods such as femtosecond transient absorption (TA) and high resolution TRPL. Trap state passivation via monolayer TMD surface treatments can be used to drastically reduce exciton trapping rates, not only enhancing ET, but isolating non-radiative loss pathways so that they can be better understood. In essence, the result shows that the emission properties of monolayer TMDs can be engineered using high PLQE QD emitters and could also be extended to electrically gated heterostructures, where the TMD monolayer is electrically pumped. Such TMD/QD heterostructures could have applications in light emitting technologies such as displays, solid-state lighting and lasers[19,22], artificial light harvesting systems or be used to read out the state of TMD devices optically in various logic and computing applications.

**Acknowledgments**

The authors thank the Winton program for physics of sustainability for financial support. We also acknowledge funding from EPSRC grants EP/L015978/1, EP/L016087/1, EP/P027741/1 and EP/P005152/1. S.D.S acknowledges support from the Royal Society and Tata Group (UF150033). GD acknowledges the Royal Society for funding through a Newton International Fellowship and the UK Engineering and Physical Sciences Research Council under grant reference EP/R023980/1. This project has received funding from the European Research Council (ERC) under the European Union's Horizon 2020 research and innovation programme (grant agreement No 758826 and 756962)

**Author contributions**

A.O.A.T fabricated and measured the samples, analysed the data and wrote the paper. N.G. built the PLE setup. G.D. and A.B. performed time resolved PL measurements. J.X. prepared the QDs. R.P. assisted in PL data analysis. J.Y.S assisted in PLE data analysis. C.A.W performed $WS_2$ steady state absorption measurements. J.A. assisted in TRPL fitting. Z.L produced TMD graphics in TOC. All authors have contributed to the writing of the manuscript.

**Figures**

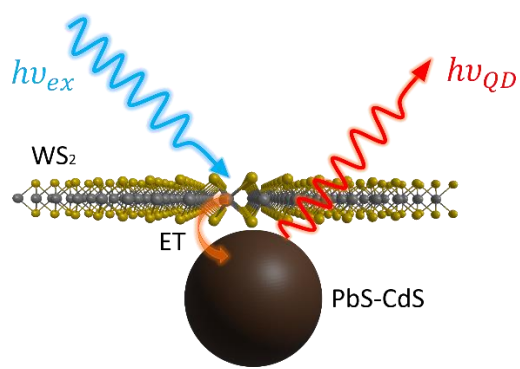

**TOC**

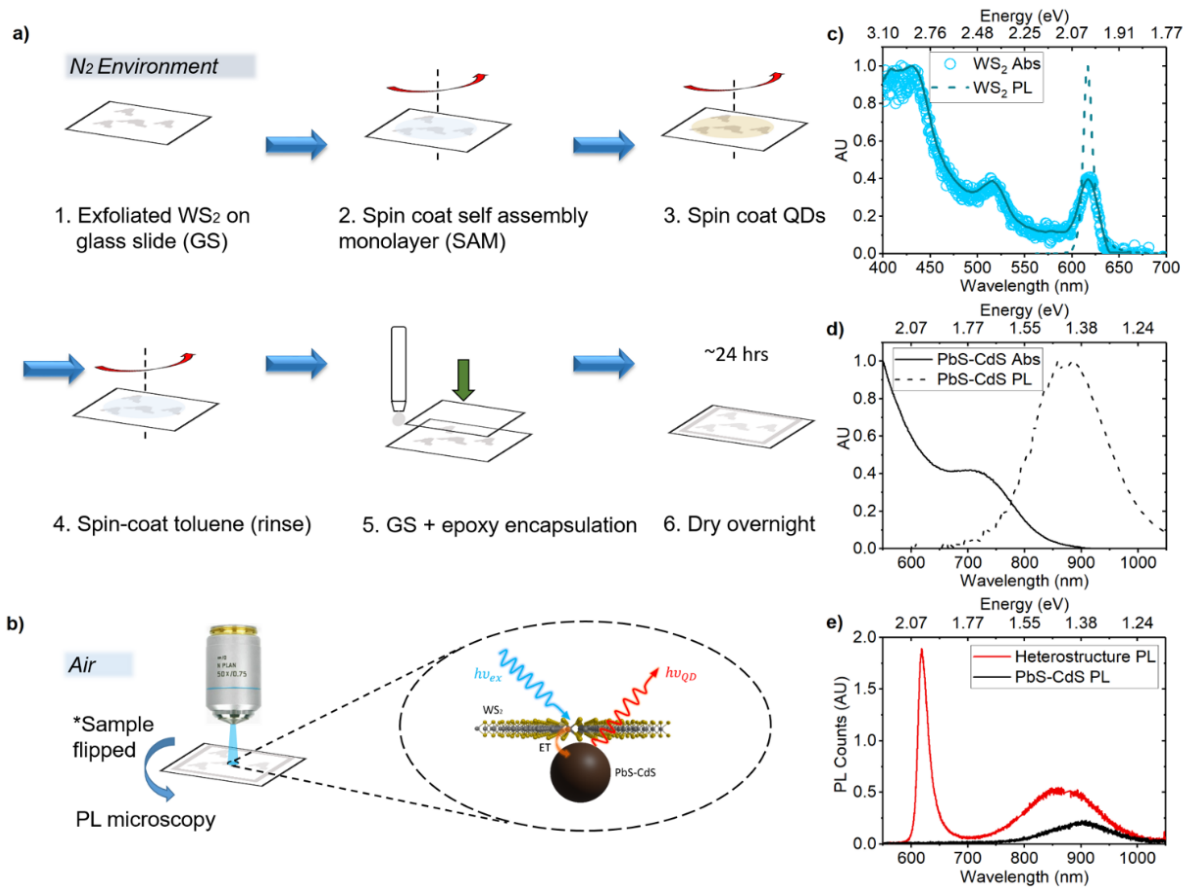

**Fig 1 a-e: a)** Cartoon illustrating heterostructure sample fabrication process (1-6) and **b)** initial PL characterization; **c)** Monolayer $WS_2$ normalised absorption (light blue circles with solid dark blue line as guide to eye) and PL (dashed dark blue line). **d)** Colloidal PbS-CdS normalised absorption (black solid line) and PL (black dashed line) spectra; **e)** PL spectra of $WS_2$-PbS-CdS 2D-QD heterostructure (red) and PbS-CdS film (black) measured with 514.5 nm CW laser at 80.2 W/cm$^2$.

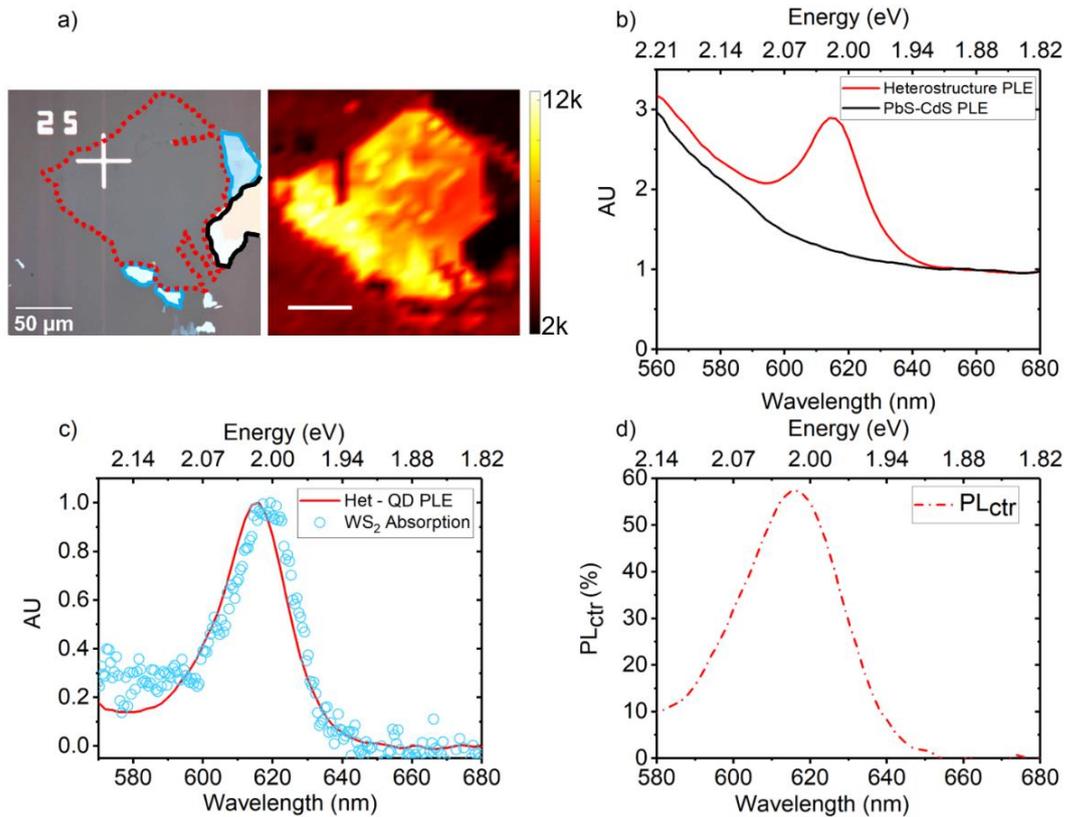

**Fig 2 a-d: a)** Optical micrograph of a WS$_2$ flake (left) showing monolayer (red dotted outline), multilayers (blue outline) and bulk crystal (black outline) with corresponding confocal NIR PL map of QD emission from the heterostructure (right) measured with 514.5 nm CW laser at 80.2 W/cm$^2$. RHS scale bar represents 50 µm; **b)** Normalised PLE spectra of heterostructure (red) and QD (control) obtained via scanning wavelengths about the WS$_2$ `A' exciton (616 nm) and detecting QD PL (900 nm). PLE spectra normalised by the average signal between 670 nm and 700 nm; **c)** Normalised `subtract' (red) signal derived via subtraction of QD PLE signal from heterostructure PLE signal in Fig. 2.b and overlapped with typical WS$_2$ absorption spectrum (blue circles); **d)** Estimated contribution to QD PL ($PL_{ctr}$) by the WS$_2$ monolayer as a function of excitation wavelength with peak value of 58% at 616 nm (~2.0 eV).

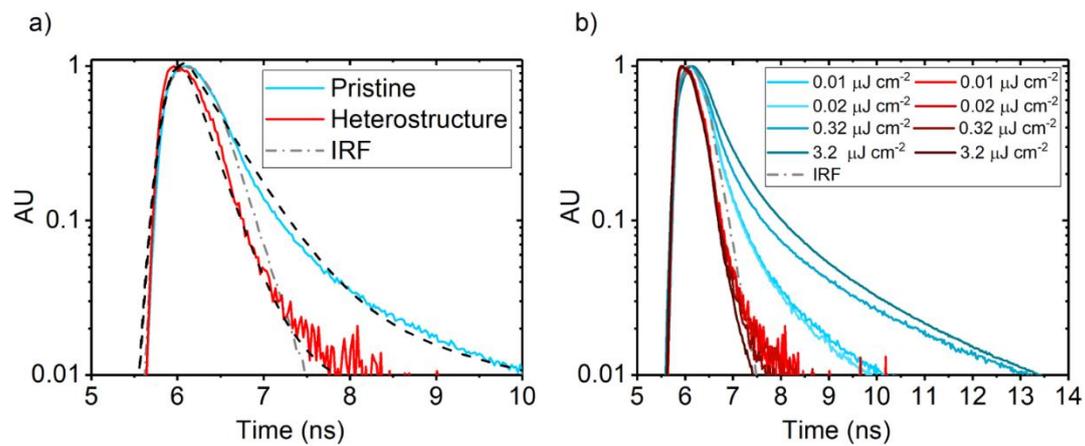

**Fig 3 a-b: a)** Low fluence time resolved $WS_2$ PL decay signals from pristine (blue) and heterostructure (red) samples measured with 509 nm pulsed excitation at 0.01 µJ/cm$^2$. Exponential decay fits are shown as dotted black lines; **b)** Time resolved $WS_2$ PL decay fluence series from pristine (blue) and heterostructure (red) samples. Pristine $WS_2$ PL decay signals show general increase in lifetime as a function of pump fluence due to exciton trapping. All $WS_2$ PL in heterostructure signal quenched below instrument response function (IRF) (grey dash-dot line) due to fast exciton transfer.

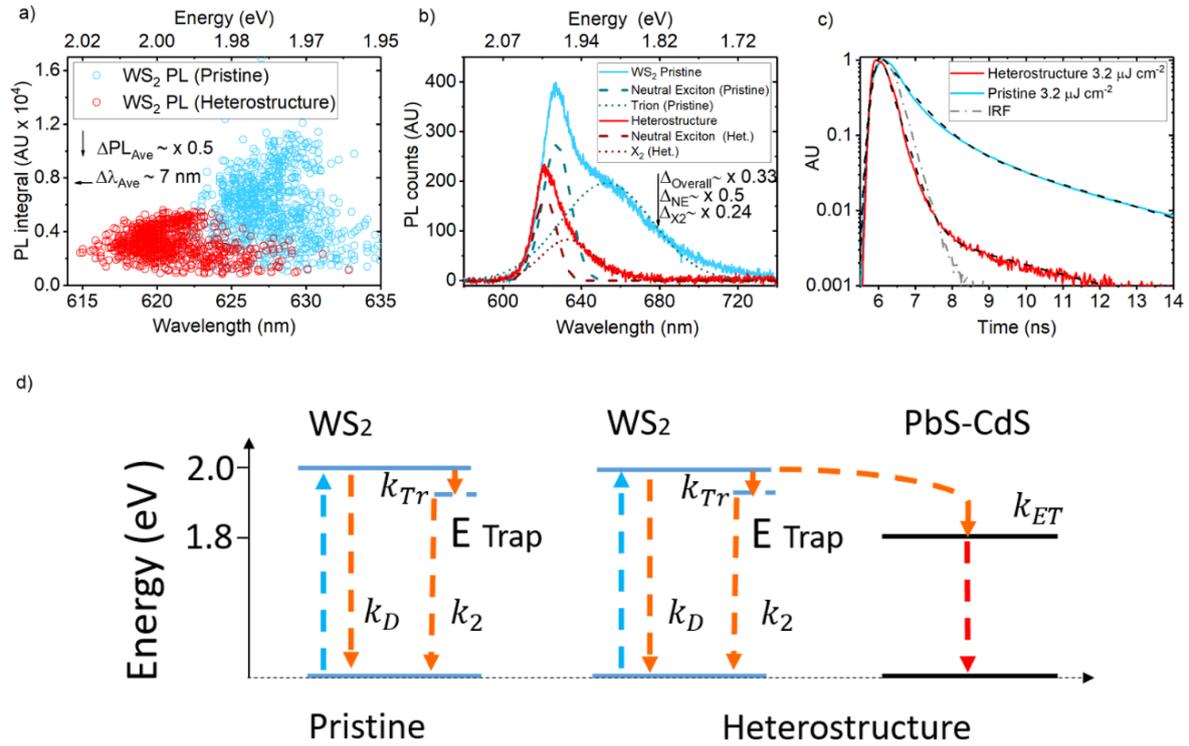

**Fig 4 a-d: a)** Scatter plot of WS$_2$ PL integrals with spectral position obtained from non-fitted PL maps of the monolayer in pristine (blue) and heterostructure (red) cases, measured with 514 nm continuous wave (CW) laser excitation at 80.2 W cm$^{-2}$ intensity; **b)** WS$_2$ PL spectra of an exemplary on the monolayer in pristine (blue) and heterostructure (red) cases. Spectra are deconvoluted with Gaussian peaks which represent the neutral exciton (dashed lines) and a lower energy species X$_2$ (dotted lines); **c)** TRPL decay spectra of pristine (blue) and heterostructure (red), measured with 509 nm excitation at 63.4 W cm$^{-2}$ intensity. Black dashed lines represent decay fits. IRF given by grey dot-dash line; **d)** Energy level diagram illustrating radiative exciton pathways in pristine WS$_2$ (LHS) and in heterostructure. Blue arrows represent initial excitation, orange arrows represent WS$_2$ excitons and red arrows represent down-shifted excitons that recombine at lower energy in the PbS-CdS QD.

# Supporting Information

# Directed Energy Transfer from Monolayer WS2 to NIR Quantum Dots

Arelo O.A Tanoh[1,2], Nicolas Gauriot[1], Géraud Delport[1], James Xiao[1], Raj Pandya[1], Joo Young Sung[1], Jesse Allardice[1], Zhaojun Li[1], Cyan A. Williams [2,3], Alan Baldwin[1], Samuel D. Stranks[1,4], Akshay Rao[1*]

[1]Cavendish Laboratory, Cambridge, JJ Thomson Avenue, CB3 0HE, Cambridge, United Kingdom

[2]Cambridge Graphene Centre, University of Cambridge, 9 JJ Thomson Avenue, Cambridge, CB3 0FA, Cambridge, United Kingdom

[3]Department of Chemistry, University of Cambridge, Lensfield Rd, CB2 1EW, Cambridge, United Kingdom

[4]Department of Chemical Engineering & Biotechnology Department of Chemical Engineering & Biotechnology, University of Cambridge, Philippa Fawcett Drive, Cambridge CB3 0AS, UK.
*E-mail: ar525@cam.ac.uk

**Section 1: Experimental Methods**

**Section 1.1: Sample Preparation**

<u>Monolayer Preparation</u>

Thin 22 mm x 22 mm glass cover slides of thickness 170 µm were solvent processed via sonication in acetone and isopropyl alcohol (IPA) for 15 mins, dried with a nitrogen ($N_2$) gun and treated in oxygen ($O_2$) plasma to remove adsorbants. Large area $WS_2$ monolayers were prepared via gold-mediated exfoliation[1]. The bulk crystal was purchased from 2D Semiconductors and exfoliated manually onto low adhesion clean-room tape prior to depositing a thin gold layer (~100-150 nm) via thermal evaporation under vacuum conditions. Once gold was evaporated, thermal release tape was adhered atop the gold coated $WS_2$ and pealed, leaving exfoliated $WS_2$ on top of a layer of gold attached to the thermal release tape. With the $WS_2$ exfoliate facing downwards, the thermal tape was affixed to the target substrate and heated on a hot plate up to 125 °C. Once the thermal tape pealed leaving the $WS_2$ exfoliate sandwiched between the substrate and gold, the excess gold was removed by gently swirling the sample immersed in potassium iodide ($KI_2$) and iodine ($I_2$) standard gold for 6 minutes. Finally, the sample was rinsed in deionised water, then sonicated in acetone for 10 mins and rinsed in IPA for 5 mins. Samples were dried with a nitrogen ($N_2$) gun. Monolayers were identified using optical contrast method[2].

<u>PbS-CdS QD preparation</u>

All chemicals were purchased from Sigma Aldrich or Romil and were used as received. The synthesis of PbS QDs was carried out following modified versions of the method of Hines & Scholes.[3]

Lead oxide (0.625 g, 99.999%), oleic acid (OA, 2 mL, 90%) and 1-octadecene (ODE, 25 mL, 90%) were placed in a three-necked round bottomed flask and degassed under vacuum at 110 °C for 2 hours with stirring, forming a colourless solution. Subsequently, the flask was put under nitrogen flow and heated to 80 °C. In a nitrogen glovebox, a syringe was prepared containing ODE (13.9 mL) and bis(trimethylsilyl)sulfide ($TMS_2S$, 296 µL, 95%). The syringe containing the sulfur precursor was rapidly injected into the reaction flask, which was allowed to cool. Upon cooling to 60 °C, the reaction mixture was transferred to an argon glovebox. The synthesised nanocrystals purified four times by precipitation with ethanol/1-butanol and acetone, centrifugation (10000 $g$) and resuspension in hexane/toluene. The purified QDs were redispersed in toluene for storage in an argon glovebox.

Cation exchange of PbS QDs was performed following a modified method of Neo et al.[4] A typical procedure was as follows:

Cadmium oxide (1.03 g, 99.999%), OA (6.35 mL) and ODE (25 mL) was placed in a three-necked round bottomed flask and degassed under vacuum for 110 °C. The vessel was switched to nitrogen and heated to 230 °C for 2 hours, resulting in the formation of a colourless solution of cadmium oleate. The solution was cooled and degassed under vacuum for 15 minutes. The flask was switched to nitrogen and the solution was transferred to a nitrogen glovebox for storage. The cadmium oleate precipitated at room temperature and was heated to 100 °C before use.

Cation exchange was performed with the addition of Cd-oleate solution to PbS nanocrystals. A typical reaction is as follows. In a nitrogen glovebox, a suspension of PbS nanocrystals in toluene (50 mg, 50 mg mL$^{-1}$) was heated to 100 °C. Cadmium oleate in ODE (0.35 mL, 0.26 M) was added to the nanocrystal suspension and maintained at 100 °C. The reaction was quenched after 1 minute with the addition of anhydrous acetone. The cation-exchanged nanocrystals were twice precipitated, centrifuged and re-suspended with acetone and toluene.

Heterostructure Preparation

Heterostructures were prepared using the following steps: In a nitrogen ($N_2$) glovebox, the monolayers on substrate were spin coated at 1000 rpm for 50 seconds with 200 µL of 20 mM 1,3 benzene dithiol dissolved in acetonitrile forming a thin self-assembly monolayer (SAM); 200 µL of 0.5 mg/ml PbS-CdS QDs ,with Oleic Acid surface attachment ligands suspended in toluene were deposited via spin coating at 500 rpm for 60 s; excess material was rinsed off by spin coating toluene on the sample at 500 rpm for 60 s. A waiting time of 5 minutes was observed between steps. Finally, the sample was encapsulated using a top 18 mm x 18 mm thin glass slide with double sided tape at the edges to hold the top slide in place. Gaps between the bottom and top glass slides were sealed with epoxy and left to dry over 24 hours in the $N_2$ environment.

It must be noted that the optical characterization (PL, PLE and TRPL) results presented herein are based on the same monolayer in pristine and heterostructure form i.e. each measurement was performed before and after QD deposition.

**Section 1.2: Optical Characterization**

Steady State Absorption and PL Spectroscopy

The absorption spectrum of the QDs was measured using a Shimadzu UV-VIS spectrometer. 0.1 mg/ml solution of colloidal QDs in toluene in a 1 cm cuvette was placed in an integrating sphere. A 1 cm cuvette filled with toluene was used as a reference. Steady state QD PL in Figure 1 c was obtained using a fluororemter (Edinburgh Instruments), with 0.1 mg/ml solution deposited in a 1 mm cuvette. Excitation wavelength was set to 500 nm and PL was detected with an indium gallium arsenide (InGaAS) array.

Steady state Absorption Microscopy

The absorption spectrum of monolayer $WS_2$ on quartz substrate was measured with a Zeiss axiovert inverted microscope in transmission using a halogen white light source via Zeiss EC Epiplan Apochromat 50x objective (numerical aperture (NA) = 0.95) forming a wide-field collection area diameter of 10 µm. Light transmitted via the sample was split with a beam splitter, with one component directed to a CCD camera (DCC3240C, Thorlabs) and the other coupled to a UV 600 nm optical fibre (200-800 nm spectral range) connected to a spectrometer (Avaspec-HS2048, Avantes).

Steady state PL microscopy

PL microscopy was performed using a Renishaw Invia confocal setup equipped with motorized piezo stage. Laser excitation was from an air-cooled Ar-ion (Argon ion) 514.5 nm continuous wave (CW) laser via 50x objective (NA = 0.75). The sample was excited upside down to ensure that the monolayer was excited first via the thin substrate to avoid shadowing by the QDs once deposited. Signals were collected in reflection via notch filter. The diffraction limited beam spot size was estimated as 0.84 µm. PL signal was dispersed via 600 l/mm grating prior to detection with inbuilt CCD detector. Laser power was measured directly via 5x objective with a Thorlabs S130C photodiode and PM100D power meter.

The detection wavelength range for PL measurements were selected using the setup's inbuilt WIRE software. The Vis-NIR PL spectrum (Figure 1.b) was generated with 10 s integration at a single spot on the heterostructure. The corresponding QD PL spectrum was taken at a location away from the heterostructure. The NIR PL map (Figure 2.a.) was generated with 8 µm resolution and 2 s integration. The Vis PL maps (Figure 4.a) was generated with 2 µm resolution and 0.5 s integration. All PL measurements were performed at 0.44 µW (80.2 W/cm$^2$).

Excitonic species were deconvoluted from pristine and heterostructure PL spectra using a procedure written in Matlab. The code incorporates the `gauss2' two Gaussian model fit. Further information on the Gaussian model is available via the *mathsworks* website.

Photoluminescence Excitation microscopy

PLE measurements were performed using a custom built inverted PL microscope setup. The inverted microscope arrangement enabled excitation of WS$_2$ monolayer first via the thin glass slide hence avoiding shadowing by the QDs. Variable wavelength excitation was provided by a pulsed super continuum white light source (Fianium Whitelase) via a Bentham TMc 300 monochromator. The optical image of the heterostructure was acquired using 600 nm laser light at low power via 60x oil objective, producing a 200 µm circular wide field image on an EMCCD camera (Photometrics QuantEM$^{TM}$ 512SC). A QD PL image of the heterostructure was obtained by filtering out the excitation wavelengths using a combination of 750 nm and 800 nm long pass filters. Further precaution was taken to remove any long wave component in the excitation line using a 750 nm short pass filter. An example of the QD PL image is given in SI figure 4, which was recorded using 620 nm excitation at 10 MHz pulse rate (~0.006 µJ/cm$^2$ fluence) and 20 s integration time. The region of interest was isolated by closing an iris in the detection line just before the camera.

The procedure for obtaining PLE spectra are as follows: i) The laser excitation via the monochromator was swept between the visible and NIR range. Given that the optics in the system were optimized for 600 nm and above, excitation was varied between 580 nm and 710 nm with 2 nm resolution. Each excitation was integrated for 20 s using 10 MHz pulses; ii) the wide field PL signal at each excitation was recorded, producing a spectrum of raw PL signal as a function of excitation wavelength; iii) the background signal was obtained by covering the detector and repeating i)-ii). The excitation power was recorded simultaneously using a Thorlabs S130C photodiode placed in the excitation line just before the sample, and a PM100D power meter interfaced with the data logging software and; iv) raw data was post-processed in *Origin* where the background spectrum was subtracted from the raw PL spectrum and normalised by the number of photons injected at each wavelength. Finally, the PLE spectrum was corrected with a system calibration file based on the PLE and absorption spectra of a high PLQE NIR dye.

Time Resolved PL microscopy

TRPL measurements were performed using a PicoQuant Microtime 200 confocal setup using a 509 nm pulsed laser excitation via an inverted 20x air objective (NA = 0.4), with estimated diffraction limited spot size of 1.55 µm. The repetition rate was set to 20 MHz with 25 ps resolution to obtain PL decay data. Signals were detected with a single photon avalanche diode (SPAD). Laser excitation was filtered out with a 510 nm long pass, and the NIR region of both pristine and heterostructure PL were filtered out using a 700 nm short pass filter, allowing for collection of WS$_2$ PL only. All signals were scaled up to 1500 s, which was used on the lowest fluence measurement in the fluence series. Power was measured using an inbuilt photodetector at each fluence, which was previously calibrated in the same experimental conditions using a standard external power-meter. Care was taken to ensure that measurements were made on the same spot on the monolayer before and after QD deposition. The instrument response function was measured with a blank glass cover slide as used for the sample.

Decay rates were fitted using a model developed in Origin, which consists of a Gaussian (as the IRF) convoluted with a double exponential decay.

**TOC Graphics**

$WS_2$ nanocrystal graphics were developed in VESTA software[5] and parsed into ChemDraw3D (Perkin Elmer) for rendering. QD graphics were modelled using Blender 3D modelling software.

**Section 2: PL contribution (PL$_{ctr}$) derivation**

The photoluminescence contribution (PL$_{ctr}$) by the TMD monolayer to the QD emitter is derived. Vavilov's rule[6], which states that PLQE is independent of excitation wavelength, forms the key assumption in this derivation. Given the range of wavelengths used in PLE measurements (580 nm – 680 nm) this assumption is regarded as reasonable. We consider the photoluminescence excitation (PLE) of the QD at excitation resonant and non-resonant to the underlying WS$_2$ monolayer i.e. $PLE_{\lambda*}$ and $PLE_{\lambda}$ respectively. In each case the PLE from the QD emission detection is given by *equations 1* and *2*:

$$PLE_{\lambda*} = \frac{PL_{\lambda*}}{n_{\lambda*}} = Abs_{\lambda*} \times PLQE \tag{1}$$

$$PLE_{\lambda} = \frac{PL_{\lambda}}{n_{\lambda}} = Abs_{\lambda} \times PLQE \tag{2}$$

Where *n* and *Abs* are the number of photons per second injected and absorption of the QDs. By dividing *equation 1* by *equation 2* we obtain:

$$\frac{PLE_{\lambda*}}{PLE_{\lambda}} = \left(\frac{n_{\lambda}}{n_{\lambda*}}\right)\frac{PL_{\lambda*}}{PL_{\lambda}} = \left(\frac{Abs_{\lambda*}}{Abs_{\lambda}}\right) \tag{3}$$

Hence the absorption ratio is equivalent to the ratio of WS$_2$ resonant PLE to non-resonant PLE. This is ratio is given by *R*:

$$R = \frac{PLE_{\lambda*}}{PLE_{\lambda}} = \left(\frac{n_{\lambda}}{n_{\lambda*}}\right)\frac{PL_{\lambda*}}{PL_{\lambda}} = \left(\frac{Abs_{\lambda*}}{Abs_{\lambda}}\right) \tag{4}$$

By comparing the *R* values on the heterostructure and the QD control, we can identify an additional contribution to the QD absorption i.e. *ΔR* from the underlying WS$_2$.

$$\Delta R = R_{Het} - R_{QD} \tag{5}$$

Expressing equation (6) as a proportion of the heterostructure R value ($R_{Het}$), we obtain the contribution of PL by the WS$_2$ to the QDs.

$$PL_{ctr} = \left(\frac{R_{Het} - R_{QD}}{R_{Het}}\right) = \left(\frac{\Delta R}{R_{Het}}\right) \tag{6}$$

**Section 3: Corroborating FRET**

Section 3.1: PLE study on heterostructures with alternative QD surface ligands of increasing length

SI figure 1 shows the wide field PLE spectra of $WS_2$/PbS-CdS heterostructures with varying QD-2D surface attachment thiol ligands. Table 1 lists the ligands used and their corresponding lengths. The difference in prominence of the $WS_2$ resonant peak is due to the variation in size of the $WS_2$ monolayers used. The BDT sample has the largest monolayer and hence the most prominent $WS_2$ `A' exciton signal with less contribution of the QD emission shoulder blue of the $WS_2$ peak as seen in other samples. All signals were obtained by scanning about $WS_2$ `A' exciton and detecting and PbS-CdS emission (~ 900 nm). All signals show the $WS_2$ `A' exciton peak in the expected spectral region (614 – 620 nm).

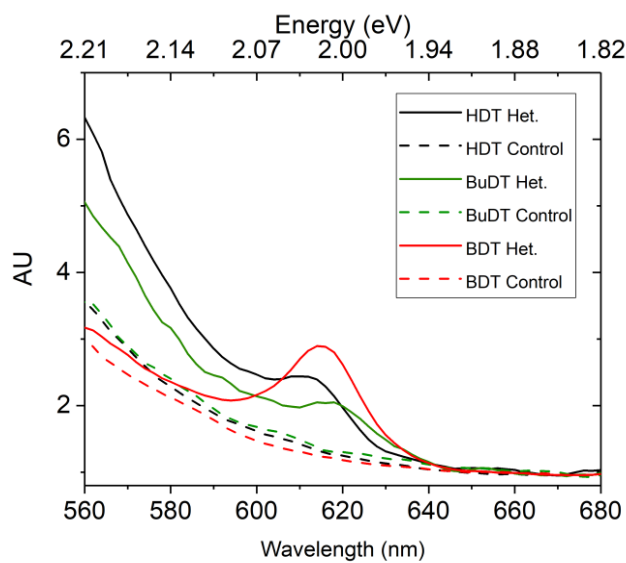

SI Fig 1: Normalised PLE spectra of 2D-QD heterostructures with varying surface ligands, namely BDT (red), BuDT (green) and HDT (black). Dashed lines represent control signals.

**Table 1**: Dithiol ligands and corresponding lengths. *(*)* indicates calculated value from known bond lengths and bond angles[7]

| Ligand | Length | Reference |
|---|---|---|
| 1,3 Benzene dithiol (BDT) | 0.53 nm | * |
| 1,4 Butane dithiol (BuDT) | 0.68 nm | 8 |
| 1,6 Hexane dithiol (HDT) | 0.95 nm | 8 |

From table 1, we note that all ligand lengths lie within range for a tunnelling energy transfer (ET) process (<1 nm) between donor and acceptor as per the requirement for Dexter energy transfer [9]. While charge orbital overlap between donor and acceptor species is a possibility at these separation distances, the high oscillator strengths of the TMD donor and QD acceptor still render ET via dipole interaction (i.e FRET) highly favourable[10], and more significant than proximity dependent DET.

The Förster radius, $R_0$, is defined as the distance between donor and acceptor through which there is a 50% probability excitation transfer[9]. we therefore estimate the theoretical Förster radius to quantify the likelihood of FRET being the dominant ET mechanism from 2D → QD in the heterostructures measured.

Section 3.2: Theoretical FRET radius estimation

Considering the 2D TMD as an array of point-like emitters and the QD film as an array of point-like absorbers, the FRET radius, $R_0$, is defined in equation 7 [10]. This system is also well approximated by a 2D quantum well donor and nanoparticle acceptors, which follows a $d^{-6}$ distance dependence for non-radiative energy transfer [11].

$$R_0^6 = \frac{9 ln 10}{128 \pi^5 N_A} \frac{\kappa^2 PLQE_D}{n^4} J \qquad (7)$$

$N_A$ is Avogadro's number, $n$ is the refractive index of the medium surrounding the FRET pair, $PLQE_D$ is the donor's intrinsic photoluminescence quantum efficiency and $\kappa^2$ is the dipole orientation factor, which is equal to 2/3 for randomly oriented dipoles[12]. $J$ is the overlap integral between the area normalised emission spectrum[10], $F_D$ and acceptor absorption spectrum given by the acceptor molar extinction coefficient, $\varepsilon_A$.

$$J = \int_0^\infty F_D(\lambda) \varepsilon_A(\lambda) \lambda^4 \, d\lambda \qquad (8)$$

It must be noted that $J$ is evaluated with the wavelength in [nm] and $\varepsilon_A$ in [M$^{-1}$ cm$^{-1}$]. To compute $R_0$, we must calculate the overlap integral $J$ from measured $\varepsilon_A(\lambda)$ and $F_D(\lambda)$ data. The molar extinction coefficient is obtained via Beer Lambert's law (*equation 9*) for absorbance, $A$ of a 0.1 mg ml$^{-1}$ suspension of QDs in toluene of molar concentration $c$, measured with a 1 cm path length, $l$, cuvette.

$$A(\lambda) = \varepsilon_A(\lambda) c l \qquad (9)$$

However, to obtain the molar extinction coefficient, the molar concentration, $c$ of QDs in [M] is needed. The first step in calculating $c$ involves estimating the QD size by solving *equation 10* provided by Moreels et al.[13] for PbS QDs of diameter $D$ using their band gap energy, $E_O$. Since the QDs used consist mainly of a PbS core as per the modified preparation method originally developed by Neo et al.[4], the use of equation 10 is considered reasonable. For the QDs used, where $E_O$ ~1.76 eV, we get $D$ ~ 2.4 nm.

$$E_0 = 0.41 + \frac{1}{0.0252 D^2 + 0.283 D} \qquad (10)$$

We then calculate the QD volume assuming a spherical shape. This is followed by multiplying the volume by the density of PbS (7.6 g cm$^{-3}$) to obtain the mass of a single QD. Multiplying the mass of a single QD by the Avogadro number yields an estimate for the QD molar mass, $Mr$ ~ 33128 g mol$^{-1}$. Dividing the known QD concentration of 0.1 g L$^{-1}$ (ie 0.1 mg mL$^{-1}$) by the estimated QD molar mass $Mr$, yields $c$ ~ 3.02 × 10$^{-6}$ M. We rearrange *equation 8* for molar extinction coefficient in [M$^{-1}$ cm$^{-1}$], which is shown in SI figure 2 along with area normalised donor emission, $F_D$.

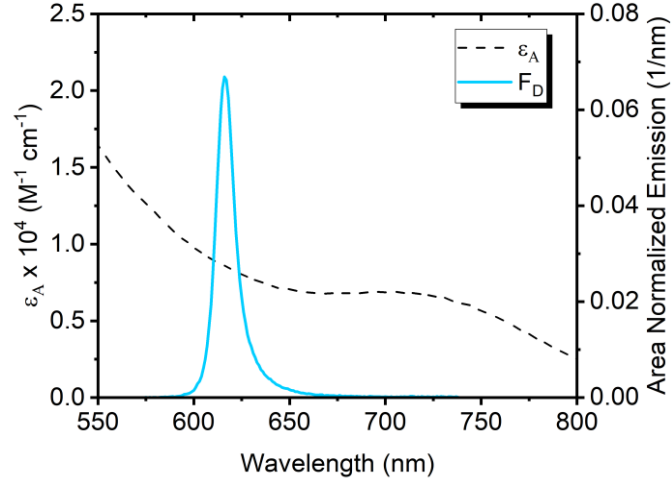

SI Fig 2: (Left axis) Molar extinction coefficient of 3.02 × 10$^{-6}$ M PbS-CdS QDs in toluene measured with 1 cm cuvette. (Right axis) Area normalized WS$_2$ emission spectrum, $F_D$ used to calculate overlap integral, $J$.

From the data shown in SI figure 2, the overlap integral is estimated via *equation 8* as $J \approx 1.23 \times 10^{15}$ M$^{-1}$ cm$^{-1}$ nm$^4$. Using a simplified version of *equation 7* below (*equation 11*) we estimate $R_0$ [nm] by assuming a vacuum between the emitter and absorber, i.e. $n$ = 1 and orientation factor $\kappa^2$ = 2/3. For the ideal system, we assume the TMD donor to have unity PLQE. This approximation is however considered reasonable as we subsequently find that the energy transfer rate from WS$_2$ band edge to QD band edge outcompetes the intrinsic exciton quenching in WS$_2$, which is the known cause of low PLQE in newly prepared TMDs.

$$R_0 = 0.0211 \left( \frac{\kappa^2 PLQE_D}{n^4} J \right)^{\frac{1}{6}} \qquad (11)$$

From *equation* 11, we obtain $\boldsymbol{R_0 \approx 6.5}$ **nm**, which exceeds the ligand separation distances between donor TMD and acceptor QD listed in table 1. This highlights the significance of the oscillator strength of the constituent heterostructure materials over their physical separation distance. This strongly implies FRET as the dominant ET process observed.

**Section 4: Energy Transfer efficiency derivation**

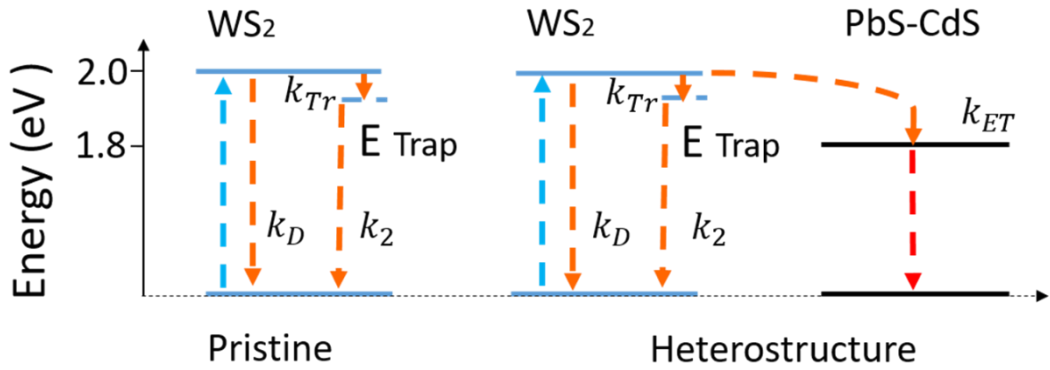

SI Fig 3: Energy level diagram illustrating radiative exciton pathways in pristine WS$_2$ (LHS) and in heterostructure. Blue arrows represent initial excitation, orange arrows represent WS$_2$ excitons and red arrows represent down-shifted excitons that recombine at lower energy in the PbS-CdS QD.

Table 2: Fitted PL lifetimes of pristine and heterostructure samples and resulting estimates for ET efficiencies. Fast components of WS$_2$ PL decay in heterostructure $\tau_1'$ and transfer efficiencies $\eta_{ET}$ represent upper bound values due to limitations in instrument sensitivity.

| Intensity | Pristine $\tau_1$ | Heterostructure $\tau_1'$ | Pristine $\tau_2$ | Heterostructure $\tau_2'$ |
|---|---|---|---|---|
| 0.21 W cm$^{-2}$ | 0.456 ns | 0.26 ns | 3.63 ns | 3.64 ns |
| 63.4 W cm$^{-2}$ | 0.62 ns | 0.26 ns | 2.95 ns | 2.9 ns |

Following the RHS of SI Figure 3. WS$_2$ donor PL kinetics in the heterostructure can be described via the following set of related ordinary differential equations (ODEs):

$$\frac{dD^*}{dt} = -(k_D + k_{TR} + k_{ET})D^* \quad (12)$$

$$\frac{dTr}{dt} = k_{TR}D^* - k_2 Tr \quad (13)$$

Where *D\** and *Tr* represent the WS$_2$ donor and trap state exciton populations respectively. The constants $k_D$, $k_{TR}$, $k_{ET}$ and $k_2$ represent the donor's intrinsic recombination rate; intrinsic trapping rate; donor-acceptor transfer rate; and trap-ground state recombination rate respectively. By integration we arrive at the solutions to *equations 12* and *13*.

$$D^*(t) = D_0^* e^{-(k_D + k_{TR} + k_{ET})t} \quad (14)$$

$$Tr(t) = \frac{k_{TR}D_0^*}{[k_2 - (k_D + k_{TR} + k_{ET})]}\left(e^{-(k_D+k_{TR}+k_{ET})t} - e^{-k_2 t}\right) \qquad (15)$$

Where $D^*_0$ represents the initial donor population. As such, the PL dynamics in the heterostructure can be defined as the sum of donor and trap population decay terms given by equations *14* and *15*:

$$PL(t) = D^*(t) + Tr(t) \qquad (16)$$

i.e

$$PL(t) = D_0^* e^{-(k_D+k_{TR}+k_{ET})t} + \left(\frac{k_{TR}D_0^*}{[k_2 - (k_D + k_{TR} + k_{ET})]}\left(e^{-(k_D+k_{TR}+k_{ET})t} - e^{-k_2 t}\right)\right) \qquad (17)$$

In the absence of the QD acceptor the pristine WS$_2$ kinetics can be modelled by setting the transfer term $k_{ET} = 0$ so that:

$$PL(t) = D_0^* e^{-(k_D+k_{TR})t} + \left(\frac{k_{TR}D_0^*}{[k_2 - (k_D + k_{TR})]}\left(e^{-(k_D+k_{TR})t} - e^{-k_2 t}\right)\right) \qquad (18)$$

The PL dynamics described by equations 17 and 18 consist of fast and slow decay components. In the pristine case (equation 18), at short time, i.e. t → 0, the fast decay time is given by:

$$\tau_1 \sim {}^1\!/\!(k_D + k_{TR}) \qquad (19)$$

Similarly, in the heterostructure case (equation 17):

$$\tau_1' \sim {}^1\!/\!(k_D + k_{TR} + k_{ET}) \qquad (20)$$

and at long time i.e t → ∞, and given that the slow decay component ($\tau_2$) remains unchanged for a given fluence, the slow decay time in both pristine and heterostructure cases is given as:

$$\tau_2 \sim \tau_2' \sim {}^1\!/\!(k_2) \qquad (21)$$

From equations 19 and 20, the ET rate $k_{ET}$ is:

$$k_{ET} = (k_D + k_{TR} + k_{ET}) - (k_D + k_{TR}) \qquad (22)$$

The ET efficiency can then be determined in terms of decay rates. Using *equations 19, 20* and *22*, the ET can be simplified in terms of fast decay rates as shown in *equation 1* of the main text.

$$\eta_{ET} = \frac{k_{ET}}{(k_D + k_{TR} + k_{ET})} = 1 - \frac{\tau_1'}{\tau_1} \qquad (23)$$

**Section 5: QD PL image on TMD monolayer**

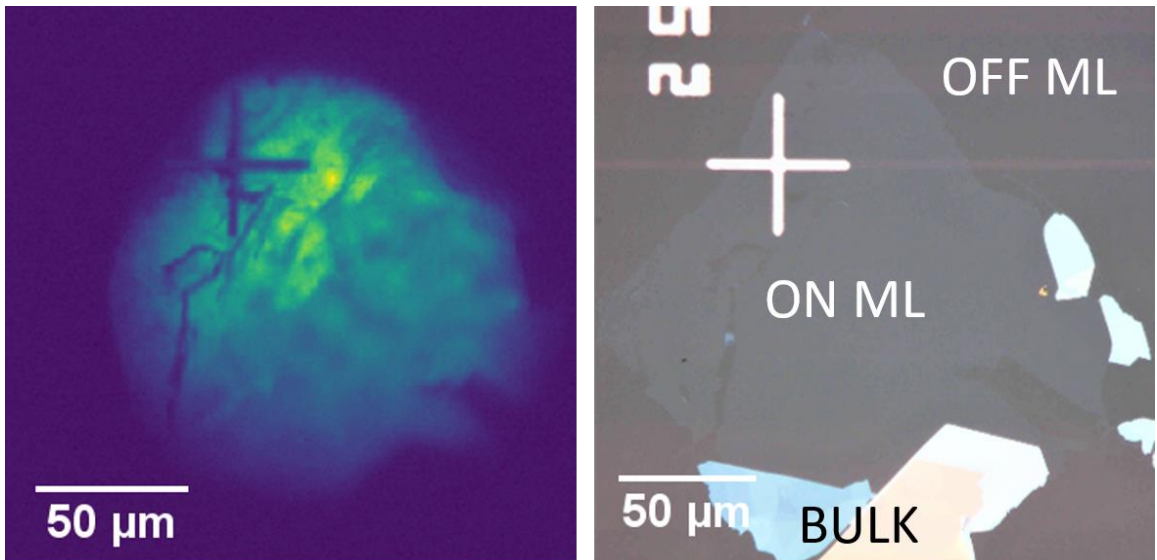

<u>SI Fig 4:</u> (LHS) Wide-field QD PL (900 nm) image of heterostructure at 620 nm excitation. (RHS) Optical image of monolayer used in heterostructure. QD PL clearly enhanced on TMD monolayer.